\documentclass[preprint, 10pt]{elsarticle}

\usepackage{amssymb}
\usepackage{amsmath}

\usepackage{dsfont}

\usepackage{url}

\usepackage{graphicx}

\usepackage[bookmarks={true}, bookmarksopen={true}, colorlinks, citecolor = blue, anchorcolor = blue,linkcolor = blue]{hyperref}
\usepackage{xcolor}
\usepackage[utf8]{inputenc}
\usepackage[T1]{fontenc}
\usepackage[english]{babel}
\usepackage{lmodern}
\usepackage{array}
\usepackage{amsfonts}
\usepackage{amsthm}
\usepackage{amsmath}
\usepackage{amssymb}
\usepackage{mathrsfs}
\usepackage{enumitem}
\usepackage{pgf, tikz}
\usepackage{bbm}
\usepackage{subfig}
\usepackage{fullpage}
\usepackage{float}
\usepackage{setspace}
\usepackage[]{natbib}

\numberwithin{equation}{section}

\renewcommand{\leq}{\leqslant}
\renewcommand{\geq}{\geqslant}

\renewcommand{\epsilon}{\varepsilon}

\newcommand{\C}{\mathbb{C}}
\newcommand{\cC}{\mathcal{C}}
\newcommand{\bC}{\mathbb{C}}

\newcommand{\E}{\mathbb{E}}

\newcommand{\N}{\mathbb{N}}
\newcommand{\R}{\mathbb{R}}

\newcommand{\der}{\mathrm{d}}

\newcommand{\defeq}{:=}

\newcommand{\arccosh}{\textnormal{arccosh}}

\newcommand{\Spec}{\textnormal{Spec}_g}

\newcommand{\com}[1]{{#1}}

\renewcommand{\omega}{{f}}

\theoremstyle{plain}
\newtheorem{prop}{Proposition}

\newtheorem{defin/prop}[prop]{Definition/Proposition}

\newlength{\myeqskip}  
\AtBeginDocument{%
    \setlength\abovedisplayskip{\myeqskip}%
    \setlength\belowdisplayskip{\myeqskip}%
    \setlength\abovedisplayshortskip{\myeqskip}%
    \setlength\belowdisplayshortskip{\myeqskip}}

\makeatletter
\def\ps@pprintTitle{%
 \let\@oddhead\@empty
 \let\@evenhead\@empty
 \def\@oddfoot{}%
 \let\@evenfoot\@oddfoot}
\makeatother

\begin{document}

\begin{frontmatter}



\title{On two fundamental properties of the zeros of spectrograms of noisy signals} 


\author[LMA]{Arnaud Poinas} 
\author[CRIStAL]{Rémi Bardenet} 

\affiliation[LMA]{organization={Université de Poitiers, LMA, UMR 7348},
            city={Poitiers},
            postcode={F-86073}, 
            country={France}}
            
\affiliation[CRIStAL]{organization={Univ. Lille, CNRS, Centrale Lille, UMR 9189 – CRIStAL},
            city={Villeneuve d’Ascq},
            postcode={59651}, 
            country={France}}

\begin{abstract}
   The spatial distribution of the zeros of the spectrogram is significantly altered when a signal is added to white Gaussian noise. 
   The zeros tend to delineate the support of the signal, and deterministic structures form in the presence of interference, as if the zeros were trapped. 
   While sophisticated methods have been proposed to detect signals as holes in the pattern of spectrogram zeros, few formal arguments have been made to support the delineation and trapping effects. 
   Through detailed computations for simple toy signals, we show that two basic mathematical arguments --the intensity of zeros and Rouché's theorem-- allow discussing delineation and trapping, and the influence of parameters like the signal-to-noise ratio.
   In particular, interfering chirps, even nearly superimposed, yield an easy-to-detect deterministic structure among zeros.
\end{abstract}


\end{frontmatter}



\begin{figure}[H]
\centering
\subfloat{\includegraphics[width=6cm]{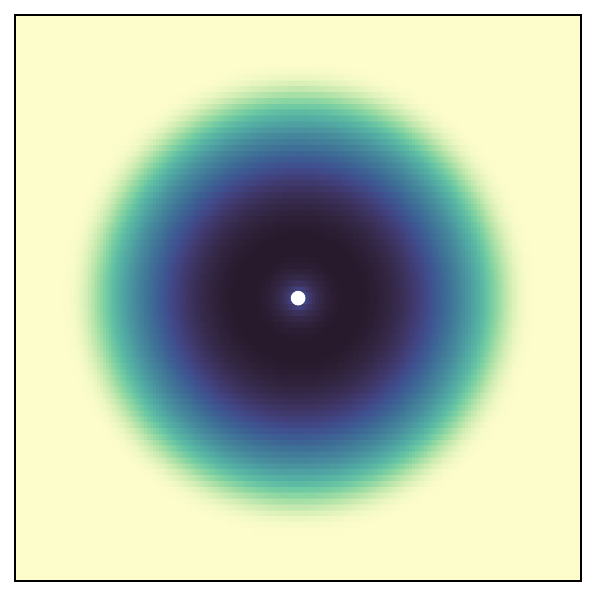}}
\subfloat{\includegraphics[width=6cm]{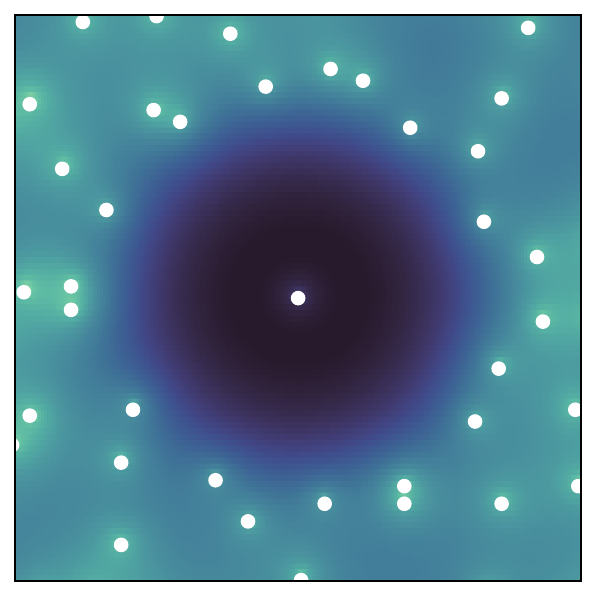}}
\caption{\small Spectrogram of an Hermite function with and without noise; zeros are white dots, low values are in yellow, large ones in dark blue.\label{fig:Example_Simple}}
\end{figure}

The short-time Fourier transform (STFT) of a signal, and its squared modulus the \emph{spectrogram}, are foundational stones of time-frequency analysis \citep{Coh95,Fla98,Gro01}. 
Classical approaches to time-frequency detection, denoising, or frequency estimation have relied on identifying regions where the spectrogram is large.
Since the seminal letter of \cite{Fla15}, \emph{zeros} of the spectrogram have also received a lot of attention.
Zeros of the spectrogram of white Gaussian noise spread very uniformly across the time-frequency (TF) plane, and \cite{Fla15} observed that, in the presence of an additional deterministic signal, the zeros tend to \emph{delineate} the support of the noiseless signal. 
Similarly, when one keeps the same realization of white Gaussian noise but increases the signal-to-noise ratio, zeros seem to become \emph{trapped} in areas surrounded by emerging signal ridges; see Figure \ref{fig:Example_Simple}.
These two foundational claims on delineation and trapping, initially based on empirical observations, have become folklore in subsequent papers, justifying e.g. equating holes in the pattern of zeros to the support of signal components \citep{BaFlCh18}.
In this letter, we give simple theoretical arguments to cement these claims, demonstrating delineation and trapping on classical examples of synthetic signals. 
In particular, we assess the effect of the signal-to-noise ratio and the relative amplitude of signal components.

In Section~\ref{s:main_results}, we give the intensity of the zeros of the STFT of a noisy signal, and a statement of Rouché's classical theorem in the context of spectrograms.
The intensity illustrates the delineation effect, while Rouché's theorem is the reason why $(i)$ zeros appear to be trapped in valleys surrounded by signal, and $(ii)$ the position of zeros crystallizes in the presence of interference.
In Section~\ref{s:specific_signals}, we apply these two results to an Hermite function and a linear chirp with added noise, while we keep the case of two interfering parallel chirps for Section~\ref{s:two_linear_chirps}. 
In particular, we illustrate the delineation and trapping effects, as well as extreme trapping resulting in near-deterministic crystallization of the zeros. 
The case of parallel chirps of different amplitudes also showcases a somewhat counterintuitive effect: informally, zeros that result from the interference between chirps move to the \emph{same} side of both chirps. 
This behavior is in stark contrast with that of the local maxima of the spectrogram, pointing to a potential advantage of zeros in the identification of interfering sources.
Finally, we provide code\footnote{\url{https://github.com/APoinas/spectrogram_zeros}} to reproduce all figures.

\textbf{Related work.} The study of the zeros of the STFT of white noise was initiated by \cite{Fla15}, and their distribution characterized in \citep{BaFlCh18, BARDENET202173}, as well as \cite{HaKoRo22} for non-Gaussian windows; see \cite{PaBa25} for a tutorial survey.
The numerical simulation of zeros is investigated in \cite{EFKR21Sub}, the distinction between zeros from white noise and zeros from signal interference in \citep{Meignen, MACLM24}.
Detection of signals based on holes can be found in several papers, from the initial \cite{Fla15} to more recent topological data analysis-based approaches \citep{MTMBG25Sub}.
We also refer to \cite{Ghosh} for a study of level sets of random spectrograms.

\section{Main results} \label{s:main_results}

Let $y:\R\mapsto\C$ be locally integrable, representing a measured signal. 
The spectrogram of $y$ with Gaussian window $g:t\mapsto 2^{1/4}e^{-\pi t^2}$ is the function
$$
   \Spec({y}): z \mapsto \left|\int_\R {y}(t){g(t-\tau)}e^{-2i\pi\omega t}\der t\right|^2
$$
of the complex variable $z=\tau+i\omega\in\C$.
The real and imaginary parts of $z$ are respectively interpreted as \emph{time} and \emph{frequency}.
The spectrogram can be rewritten as $\Spec(y)(z)=e^{-\pi|z|^2}|B(y)(\bar z)|^2$, where $B(y)$ is a holomorphic function called the \emph{Bargmann transform} of $y$; see \cite{Folland89,Gro01}. 
For the remainder of the paper, we consider $y = x + \xi$ to be the sum of a deterministic signal $x$ and a complex white Gaussian noise $\xi$; see \cite{BARDENET202173} for a detailed mathematical definition of $\xi$. 
The zeros of $\Spec(x+\xi)$ are a random configuration of points in $\mathbb{C}$, i.e. a \emph{point process}.
Our aim is to describe a few properties of that point process that relate to signal detection and reconstruction.

\subsection{Explaining delineation: the intensity of zeros}
We first wish to explain the general tendency of zeros to avoid locations where the spectrogram of the noiseless signal $x$ is large, while concentrating around that region. 
For that purpose we compute the intensity function $\rho{:\mathbb{C}\rightarrow \R_+}$ of the zeros. 
Denoting by $N(A)$ the number of zeros of the spectrogram in a bounded Borel set $A\subset\C$, the intensity function is defined as the density with respect to Lebesgue of the measure $\mathbb{E}N(\cdot)$, so that $\E[N(A)]=\int_A \rho(z)\der z$. 
Independently of our work and for a different purpose, an expression of $\rho$ has already been established in \citep{EFKR21Sub}, as a function of the Bargmann transform of the noiseless signal.
We rather use the spectrogram of the signal, which makes $\rho$ easier to interpret.

\begin{prop}\label{prop:zero_intens_signal}
Let $z\in\C$. 
Then
\begin{equation}
   \label{eq:zero_intens_signal}
   \rho(z)=\left(1 + \Spec(x)(z) + \frac{\Delta \Spec(x)(z)}{4\pi}\right)\exp\left(-\Spec(x)(z)\right),
\end{equation}
where \com{$\Delta \Spec(x)(z)$} is the Laplacian of the function $\tau, f\mapsto \Spec(x)(\tau + i f)$.
\end{prop}
\begin{proof}
By the linearity of the Bargmann transform, we have 
$$ \Spec(x+\xi)(\bar z)=e^{-\pi|z|^2}|B(x)(z)+B(\xi)(z)|^2, $$ 
so that the zeros of $z\mapsto \Spec(x+\xi)(\bar z)$ share the same distribution as the zeros of $B(x)+B(\xi)$.
The latter is a Gaussian field with mean $B(x)$ and covariance kernel $K(z, w)=e^{\pi z \bar w}$; see e.g. \cite{BaFlCh18}. 
The density of zeros of a Gaussian field can be computed with the Kac-Rice formula \cite[Theorem 6.2]{Azais} after establishing that, almost surely, none of the zeros of $B(x)+B(\xi)$ are critical points. Following a similar idea to \cite{Abreu} and \cite{Feng}, we introduce for convenience the linear operator 
\begin{equation}\label{eq: def_nabla}
\nabla_z'\defeq \partial_z - \pi \bar z,
\end{equation}
\com{where $\partial_z$ denotes the Wirtinger derivative. We refer the interested reader to \cite[Appendix 2]{Wirtinger} for the main properties of these derivatives.} 
Any critical point $z$ of $B(x)+B(\xi)$ that is also a zero satisfies
\begin{equation}
   \label{e:zero_and_critical_initial}
   (B(x)+B(\xi))(z) = \partial_z(B(x)+B(\xi))(z) = 0,
\end{equation}
\com{and thus also}
\begin{align}
   \nabla_z'(B(x)+B(\xi))(z) &= 0.
   \label{e:zero_and_critical}
\end{align}
We will show that, with probability $1$, no $z$ satisfies \ref{e:zero_and_critical}.
First, note that $\nabla_z' B(\xi)$ is a Gaussian field with kernel
\begin{equation}\label{eq: Nabla_kernel}
   \forall z, w\in\C,~\E[\nabla_z'B(\xi)(z)\overline{\nabla_w'B(\xi)(w)}]=\nabla_z'\nabla_{\bar w}'K(z, w)=\pi(1-\pi|w-z|^2)e^{\pi z\bar w}.
\end{equation}
Moreover, the covariance between $B(\xi)$ and $\nabla_z' B(\xi)$ satisfies
$$
   \forall z, w\in\C,~\E[\nabla_z'B(\xi)(z)\overline{B(\xi)(w)}]=\nabla_z'K(z, w)=\pi(\bar w-\bar z)e^{\pi z\bar w}.
$$
This covariance vanishes when $z=w$, so that, for all $z\in\C$, $\nabla_z'B(\xi)(z)$ and $B(\xi)(z)$ are independent. \com{Since the variances of both $B(\xi)$ and $\nabla_z'B(\xi)$ never vanish, the probability of any $z\in\C$ simultaneously being a zero of both $B(x)+B(\xi)$ and $\nabla_z'(B(x)+B(\xi))$, and thus of $\partial_z(B(x)+B(\xi))$, is $0$. 
Using Lemma 28 in \citep{PV05}, this implies that $B(x)+B(\xi)$ almost surely has no multiple zero.}

We can therefore use the Kac-Rice formula, yielding, for $z\in\mathbb{C}$, 
\begin{equation}\label{eq: Cond_Exp_KR}
\rho(\bar z)=\E\left[\left.\det\begin{pmatrix}
\partial_z (B(x)(z)+B(\xi)(z)) & \partial_z (\overline{B(x)(z)+B(\xi)(z))} \\
\partial_{\bar z} (B(x)(z)+B(\xi)(z)) & \partial_{\bar z} (\overline{B(x)(z)+B(\xi)(z))}
\end{pmatrix} \right| B(x)(z)+B(\xi)(z)=0\right]p_{B(x)(z)+B(\xi)(z)}(0),
\end{equation}
where $p_{B(x)(z)+B(\xi)(z)}$ is the probability density of $B(x)(z)+B(\xi)(z)$, i.e.
$$
   p_{B(x)(z)+B(\xi)(z)}(0)=\frac{1}{\pi e^{\pi|z|^2}}\exp\left(-\frac{|B(x)(z)|^2}{e^{\pi |z|^2}}\right).
$$
Now, $B(x)$ and $B(\xi)$ being holomorphic, combined with \eqref{eq: Nabla_kernel} and the independence between $B(\xi)(z)$ and $\nabla_z' B(\xi)(z)$, we obtain
\begin{align}
   \rho(\bar z)&=\frac{1}{\pi e^{\pi|z|^2}}\E\left[|\partial_z (B(x)(z)+B(\xi)(z))|^2 \,\right|\left. B(x)(z)+B(\xi)(z)=0\right]\exp\left(-\frac{|B(x)(z)|^2}{e^{\pi |z|^2}}\right)\nonumber\\
   &=\frac{1}{\pi e^{\pi|z|^2}}\E\left.\left[|\nabla_z' (B(x)(z)+B(\xi)(z))|^2 \,\right| B(x)(z)+B(\xi)(z)=0\right]\exp\left(-\frac{|B(x)(z)|^2}{e^{\pi |z|^2}}\right)\nonumber\\
   &=\frac{1}{\pi e^{\pi|z|^2}}\left(\E\left[|\nabla_{z}'B(\xi)(z)|^2\right] + |\nabla_{z}'B(x)(z)|^2\right)\exp\left(-\frac{|B(x)(z)|^2}{e^{\pi |z|^2}}\right)\nonumber.
\end{align}
In particular, 
\begin{align}
   \rho(z)&=\left(1 + \frac{|\nabla_{z}'B(x)(\bar{z})|^2}{\pi e^{\pi |z|^2}}\right)\exp\left(-\frac{|B(x)(\bar{z})|^2}{e^{\pi |z|^2}}\right), \quad z\in\mathbb{C}.\label{eq: rho_alternative}
\end{align}
One can recognize that the term inside the exponential in \eqref{eq: rho_alternative} corresponds to $-\Spec(x)(z)$. As for the other term, we have
$$\partial_z \Spec(x)(z)= \partial_z\left(e^{-\pi|z|^2}B(x)(\bar{z})\overline{B(x)(\bar{z})}\right) = \partial_z\left(e^{-\pi|z|^2}B(x)(\bar{z})B(\bar x)(z)\right)= e^{-\pi|z|^2}B(x)(\bar{z})\nabla_z'(B(\bar x)(z)).$$
Hence, $\Delta \Spec(x)(z)= 4\partial_{\bar z}\partial_z \Spec(x)(z)$ expands into
\begin{align}
4\partial_{\bar z} \left(e^{-\pi|z|^2}B(x)(\bar{z})\nabla_z'(B(\bar x)(z))\right)=~& 4e^{-\pi|z|^2}\nabla_{\bar z}'(B(x)(\bar z))\nabla_z'(B(\bar x)(z))-4\pi e^{-\pi|z|^2}B(x)(\bar{z})B(\bar x)(z)\nonumber\\
=~& 4\frac{|\nabla_{\bar z}'(B(x)(\bar{z}))|^2}{e^{\pi |z|^2}}-4\pi \Spec(x)(z).\label{eq: Positivity}
\end{align}
Combined with \eqref{eq: rho_alternative}, we obtain
\begin{equation}\label{eq: rho}
\rho(z)=\left(1 + \Spec(x)(z) + \frac{\Delta \Spec(x)(z)}{4\pi}\right)\exp\left(-\Spec(x)(z)\right), \quad z\in\mathbb{C}.
\qedhere
\end{equation}
\end{proof}

The exponential term in \eqref{eq:zero_intens_signal} explains why zeros avoid locations where the spectrogram of the signal is large. 
Now, as a \com{consequence of identity \eqref{eq: Positivity} we get that $4\pi \Spec(x)(z) + \Delta \Spec(x)(z)\geq 0$ for all $z\in\C$. So,} when $\Spec(x)(z)\approx 0$ \com{but $\Delta \Spec(x)(z)$ is not close to $0$}, $\rho(z)$ is approximately $1 + \Delta \Spec(x)(z) / 4\pi$, while $\Delta \Spec(x)(z)\geq 0$.
Consequently, zeros tends to accumulate around the contours of the signal support, defined as regions where the spectrogram of the noiseless signal $x$ is negligible but $\Delta \Spec(x)(z)$ is not. 
Finally, in locations where $\Spec(x)(z)$ is both negligible and has barely any variation, $\rho(z)\approx 1$, which is the intensity of the zeros of the spectrogram of white noise only \citep{BaFlCh18}.

\subsection{Explaining trapping: Rouché's theorem}
One specific behavior of zeros of spectrogram of noisy signals that cannot be explained by the intensity function is their tendency to get "trapped". 
As can be observed in Figures \ref{fig:Example_Simple} and \ref{fig: Two_chirps}, when the spectrogram of a signal has a zero surrounded by high values then, when noise is added to the signal, the same zero will still be present at around the same location. 
This behavior is a direct consequence of Rouché's theorem, see e.g. \cite{Rud87}
\begin{prop}
   \label{prop: Rouche}
   Let $\cC$ be a closed, simple curve of $\C$.
   Let $x:\R\rightarrow\C$ be a locally integrable function, and let $\xi$ be the white Gaussian noise introduced in Section \ref{s:main_results}. 
   On any event such that 
   \begin{equation}
      \label{e:domination}
      \forall z\in \cC,~~\Spec(\xi)(z) < \Spec(x)(z),
   \end{equation} 
   $\Spec(\xi + x)$ has the same number of zeros as $ \Spec(x)$ in the interior of $\cC$, where each zero is counted as many times as its multiplicity.
\end{prop}
\begin{proof}
We know that $\Spec(x+\xi)(z)=e^{-\pi|z|^2}\com{|B(x)(\bar z)+B(\xi)(\bar z)|^2}$.
If we denote by $\bar \cC$ the set of $\bar z$ such that $z\in \cC$, which is also a closed simple curve of $\C$, then $\Spec(x+\xi)$ has the same number of zeros in the interior of $\cC$ as $B(x)+B(\xi)$ has in the interior of $\bar \cC$. 
We showed in the proof of Proposition \ref{prop:zero_intens_signal} that all zeros of $B(x)+B(\xi)$ are almost surely of multiplicity equal to one. 
By assumption, $\Spec(\xi)(z) < \Spec(x)(z)$ for all $z\in \cC$ and thus 
$$
   \forall z\in\bar \cC,~|B(x)(z)+B(\xi)(z) - B(x)(z)| < |B(x)(z)|.
$$
Since $B(x)$ and $B(\xi)$ are holomorphic, Rouché's theorem states that $B(x)+B(\xi)$ and $B(x)$ share the same number of zeros in the interior of $\bar \cC$, where each zero of $B(x)$ is counted as many times as its multiplicity. 
The same conclusion holds for the zeros of $\Spec(x+\xi)$ and $\Spec(x)$ in the interior of $\cC$.
\end{proof}
This result show that when $\Spec(x)$ has large enough values on a closed curve $\cC$, then $\Spec(\xi + x)$ is very likely to have the same number of zeros as $\Spec(x)$ in the interior of $\cC$, counted with multiplicity. In later applications of Rouché's theorem, we shall need to control the probability that the spectrogram of the noise is dominated by the spectrogram of the noiseless signal on a given contour, as in \eqref{e:domination}. We derive a generic lower bound on that probability in \ref{sec: Gaussian_bound}.

\section{Application to specific signals} \label{s:specific_signals}

\begin{figure}[H]
   \centering
   \subfloat{\includegraphics[width=4cm]{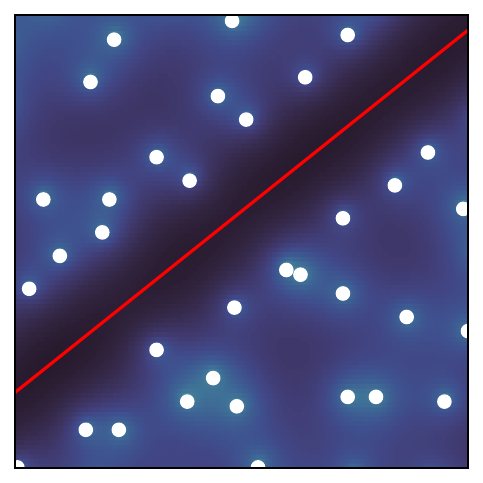}}
   \hfill
   \subfloat{\includegraphics[width=4cm]{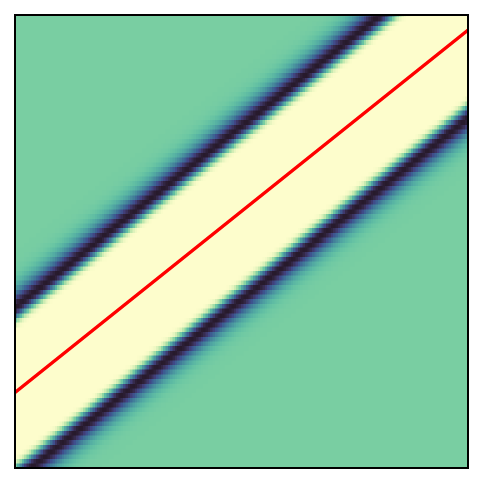}}
   \hfill
   \subfloat{\includegraphics[width=4cm]{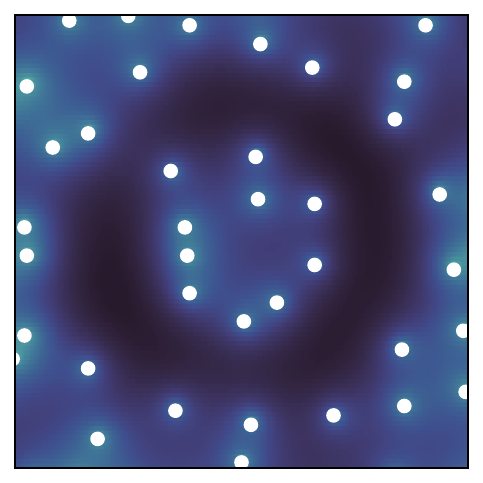}}
\hfill
\subfloat{\includegraphics[width=4cm]{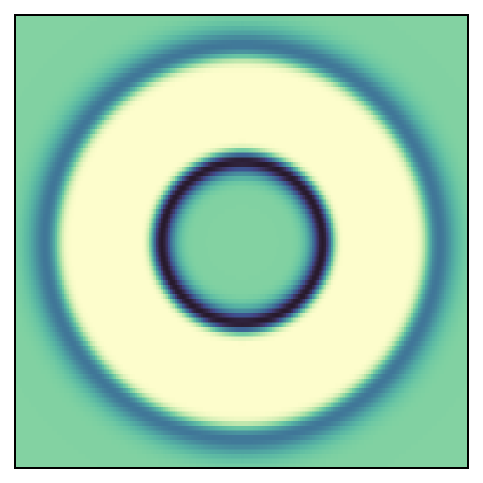}}
   \caption{\small Spectrograms of noisy signals and associated density of zeros. Left: Noisy chirp with $\gamma=100$, $a=-5$ and $b=0.4$. Right: Noisy Hermite function with $\gamma=400$ and $k=10$. \label{fig: Hermite + Single_chirp}}
\end{figure}

\subsection{Hermite functions}

Let $H_k$ be the $k$-th orthogonal polynomial with respect to $t\mapsto e^{-2\pi t^2}$. 
Let $h_k(\cdot) \propto H_k(\cdot)g(\cdot)$ be the $k$-th \emph{Hermite function}, normalized so that $\|h_k\|_2=1$. 
$h_0, h_1, \dots$ form an orthonormal basis of $\mathbb{L}^2(\R^d)$ satisfying
\begin{equation}\label{eq: Spec_hermite}
B(h_k): z \mapsto \frac{\pi^{k/2} z^{k}}{\sqrt{k!}}~\mbox{and}~\Spec(h_k): z \mapsto e^{-\pi|z|^2}\frac{\pi^k|z|^{2k}}{k!};
\end{equation}
\citep[Section 3.4]{Gro01}.
Figure~\ref{fig: Hermite + Single_chirp} shows the spectrogram of an Hermite function plus white noise. 
As can be guessed, injecting \eqref{eq: Spec_hermite} into \eqref{eq: rho_alternative} shows that the intensity of the zeros is a radial function.

\begin{prop}\label{Prop: Hermite_Density}
Let $k\in \mathbb{N}^*$ and $\gamma>0$. 
The intensity $\rho$ of zeros of $\Spec(\xi + \sqrt{\gamma} h_k)$ only depends on $r=\pi|z|^2$, and is equal to
$$
\rho(r)=\left(1 + \gamma\frac{r^{k-1}}{k!}\left(k-r\right)^2 e^{-r} \right)\exp\left(-\gamma\frac{r^k}{k!}e^{-r}\right).
$$
\end{prop}

As a consequence, we also get the average number of zeros falling into $B(0, R)$, the centered ball with radius $R$:
\begin{equation} \label{eq: Expect_ball_hermite}
\E[N(B(0,R))]=\int_{B(0,R)}\rho(\pi|z|^2)\der z=\int_{0}^{\pi R^2}\rho(r)\der r =  k - (k-\pi R^2)\exp\left(\hspace{-0.05cm}-\gamma\frac{\pi^k R^{2k}}{k!}e^{-\pi R^2}\hspace{-0.05cm}\right).
\end{equation}
When $R=\sqrt{k/\pi}$, the average number of zeros falling in $B(0, R)$ is equal to $k$ and does not depend on the \com{amplitude scaling factor} $\gamma$ \com{(hereafter, we abusively call $\gamma$ the signal-to-noise ratio, or SNR)}. 
In fact, since $B(h_k)$ has a unique zero at $z=0$ with a multiplicity of $k$ and the largest value of $\Spec(h_k)(z)$ is $k^ke^{-k}/k!$, reached when $\|z\|=\sqrt{k/\pi}$, Propositions \ref{prop: Rouche} and \ref{prop: Gaussian_bound} with the curve $\mathcal{C}$ being the circle with radius $\sqrt{k/\pi}$ yields the following result.
The number of zeros in $B(0, R)$ is exactly $k$ with probability at least $1-O(e^{-C\gamma})$ for some constant $C>0$. This is a clear illustration of the \emph{trapping} effect.

\begin{prop} \label{Prop: Hermite_Trapping}
Let $\epsilon\in (0, 1/4)$. There exists a constant $M_k>0$ only depending on $k$ such that if
$
   \gamma > 2k^{-k}e^kk!(M_k + \sqrt{\log(4/\epsilon)})^2,
$
then
$$
   \mathbb{P}\left(N\left(B\left(0,\sqrt{k/\pi}\right)\right)=k\right)\geq 1 - \epsilon.
$$
\end{prop}

\subsection{Linear chirp}

Let $a,b\in\mathbb{R}$.
We now consider a linear chirp
\begin{equation}
   \label{e:chirp}
   x:t\mapsto e^{2i\pi t(a+bt)}.
\end{equation}
Letting $\sigma_b = \sqrt{\frac{2}{1+4b^2}}$, its spectrogram is \citep{Behera}
\begin{equation}
   \label{e:spectrogram_single_chirp}
   \Spec(x)(\tau + i\omega) = \sigma_b \exp\left(-\pi\sigma_b^2(\omega-(a+2b\tau))^2 \right).
\end{equation}

\begin{prop} \label{Prop: Density_Chirp}
   Let $a,b\in\R$, $x$ be the chirp in \eqref{e:chirp}, and $\gamma>0$. 
   The density $\rho$ of the zeros of $\Spec(\xi + \sqrt{\gamma} x)$, as a function of $z=\tau + i \omega$, only depends on $r\defeq\frac{\sigma_b}{\sqrt{2}}(\omega-(a + 2b\tau))$, the signed distance between $z$ and the line $\omega = a+2b\tau$.
   Furthermore,
   $$
      \rho(r)=\left(1 + 4\pi \gamma\sigma_b r^2e^{-2\pi r^2}\right)\exp\left(-\gamma\sigma_b e^{-2\pi r^2}\right).
   $$
\end{prop}

We do not provide a proof as this result can be recovered from Proposition \ref{prop: intensity_zeros_2chirps} by taking $\gamma_2=0$. Proposition~\ref{Prop: Density_Chirp} shows how zeros will likely be absent along the chirp's axis, while the competing quadratic and negative exponential terms will draw two parallel bands along the support of the chirp where zeros will be marginally likely to occur; see also Figure~\ref{fig: Hermite + Single_chirp}.
The higher the SNR $\gamma$, the larger the distance between the two bands.
Moreover, let $\mathcal{R}_R$ be any rectangle with a side of length $1$ located on the main axis of the linear chirp and whose perpendicular sides are of length $R$. 
Explicit integration shows that 
\begin{equation}
   \label{e:expected_number}
   \E[N(\mathcal{R}_R)] = \int_0^R \rho(r)\der r = R \exp\left(-\gamma\sigma_b e^{-2\pi R^2}\right).
\end{equation}
This can be interpreted as follows: when the SNR $\gamma$ is $0$, the expected number of zeros is $R$. 
The negative exponential in \eqref{e:expected_number} thus witnesses the fraction of zeros that are "pushed" out of the rectangle as $\gamma$ grows.
As the chirp's slope $b$ grows, the "width" $\sigma_b$ of the chirp's support in \eqref{e:spectrogram_single_chirp} decreases, and so does the average number of expelled zeros.

\section{A pair of parallel linear chirps} \label{s:two_linear_chirps}

\begin{figure}[H]
\centering
\subfloat{\includegraphics[width=4cm]{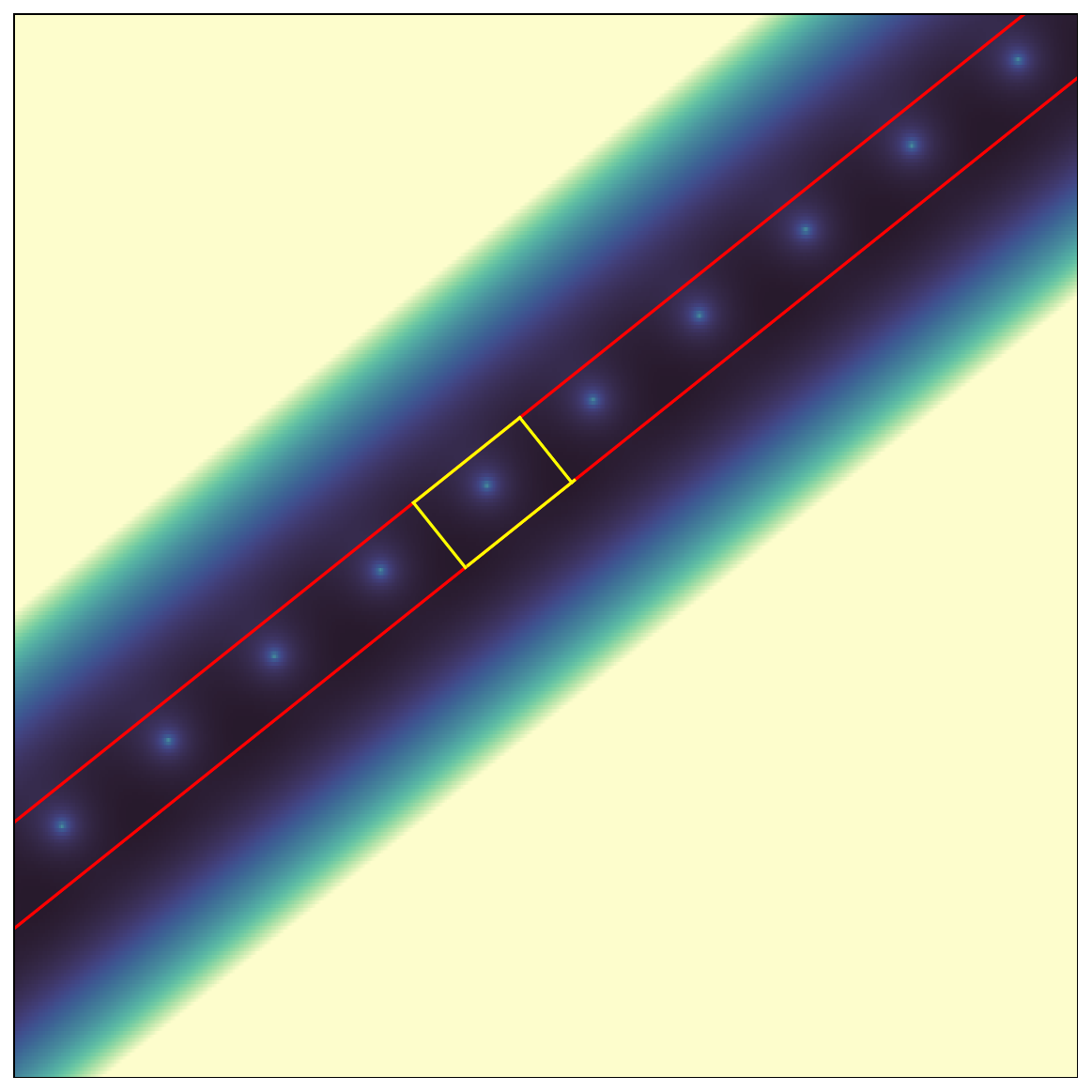}}
\hspace{1cm}
\subfloat{
\begin{tikzpicture}[scale=1.5]

\node at (1.5 + 2/13 - 0.2, 1 - 3/13 - 0.15) {$a$};
\node (P) at (0.8, 1.6) {+};
\node at (0.55, 1.75) {$z$};
\node at (0.85, 1.1) {$r$};
\node at (1.1, 2) {$s$};
\node [rotate=33.69] at (3.25, 1.3) {\scriptsize $\omega = a_2 + 2b\tau$};
\node [rotate=33.69] at (2.9, 1.75) {\scriptsize $\omega = a_1 + 2b\tau$};

\draw [red] (3*0.2, 2*0.2) -- (3*1.2, 2*1.2);
\draw [red] (1, 0) -- (4*1, 2*1);
\draw (1.5, 1) -- (1.5+4/13, 1-6/13);
\draw [dashed] (2 + 2 * 0.3, 4/3 - 3 * 0.3) -- (2 - 2 * 0.4, 4/3 + 3 * 0.4);
\draw (P.center) -- (3*5.6/13, 2*5.6/13);
\draw (P.center) -- (2-2*3.2/13, 4/3 + 3 * 3.2/13);
\end{tikzpicture}
}
\caption{\label{fig: Example_TwoChirps} \small Left: Spectrogram of a noiseless pair of chirps with $(\gamma_1, \gamma_2, a_1, a_2, b)=(100, 40, -1, 0, 0.4)$. One of the rectangles $\cC_N$ in which a zero is trapped is shown in yellow. Right: Change of coordinates.}
\end{figure}
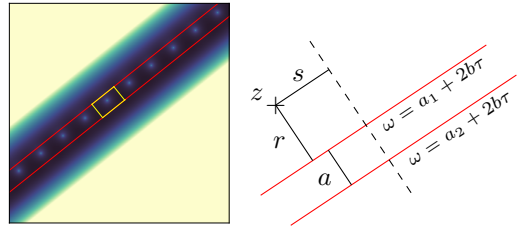

To investigate the effect of interference, we consider the superposition of two parallel linear chirps with respective amplitudes $\gamma_1, \gamma_2>0$,
\begin{equation*}
   x:t\mapsto \sqrt{\gamma_1}e^{2i\pi t(a_1+bt)}+\sqrt{\gamma_2}e^{2i\pi t(a_2+bt)}.
\end{equation*}
For $z=\tau+i\omega\in\C$, the spectrogram of $x$ is
\begin{equation}\label{eq:spectro_double_chirp}
\Spec(x)(z)=\sigma_b\left(\gamma_1e^{-2\pi r^2}+\gamma_2e^{-2\pi(r-a)^2}+2\sqrt{\gamma_1\gamma_2} e^{-\pi r^2-\pi(r-a)^2}\cos\left(2\pi as\right)\right),
\end{equation}
where $a\defeq \frac{\sigma_b}{\sqrt{2}}(a_2 - a_1)$, $r\defeq \frac{\sigma_b}{\sqrt{2}}(\omega-2b\tau-a_1)$ and $s\defeq \frac{\sigma_b}{\sqrt{2}}(\tau+2b\omega-(a_1+a_2)b)$; see Figure \ref{fig: Example_TwoChirps}. 
While a similar expression already exists in the literature \citep{Meignen} it is under a slightly different setting. For the sake of completeness, we go through a quick proof of their expression in our framework in \ref{sec: Spectro_two_chirps}. Unlike for a single chirp, the interference between the two chirps makes the spectrogram vanish in some points. Writing
$$\Spec(x)(z)\geq\sigma_b\left(\gamma_1e^{-2\pi r^2}+\gamma_2e^{-2\pi(r-a)^2}-2\sqrt{\gamma_1\gamma_2} e^{-\pi r^2-\pi(r-a)^2}\right)=\sigma_b\left(\sqrt{\gamma_1}e^{-\pi r^2}-\sqrt{\gamma_2}e^{-\pi(r-a)^2}\right)^2$$
with equality if and only if $\cos(2\pi as)=0$ we get

\begin{prop}\label{prop: location_zeros}
All zeros of $\Spec(x)$ are simple and satisfy
\begin{equation}\label{eq:location_zeros}
r=\frac{a}{2}-\frac{\log(\gamma_2/\gamma_1)}{4a\pi}~\textnormal{and}~as-\frac{1}{2}\in\mathbb{Z}.
\end{equation}
\end{prop}

The zeros of the noiseless spectrogram $\Spec(x)$ thus group on a line parallel to the axes of the chirps. 
If both chirps have the same amplitude, then this line is exactly in the middle of their axes. 
Otherwise, it is shifted towards the weakest signal. 
Interestingly, the distance between the zeros only depends on the distance between the two axes of the chirps, but is unaffected by their amplitudes.
We now look at the zeros of the noisy spectrogram $\Spec(\xi+x)$.

\begin{figure}[H]
\centering
\subfloat{\includegraphics[width=3.5cm]{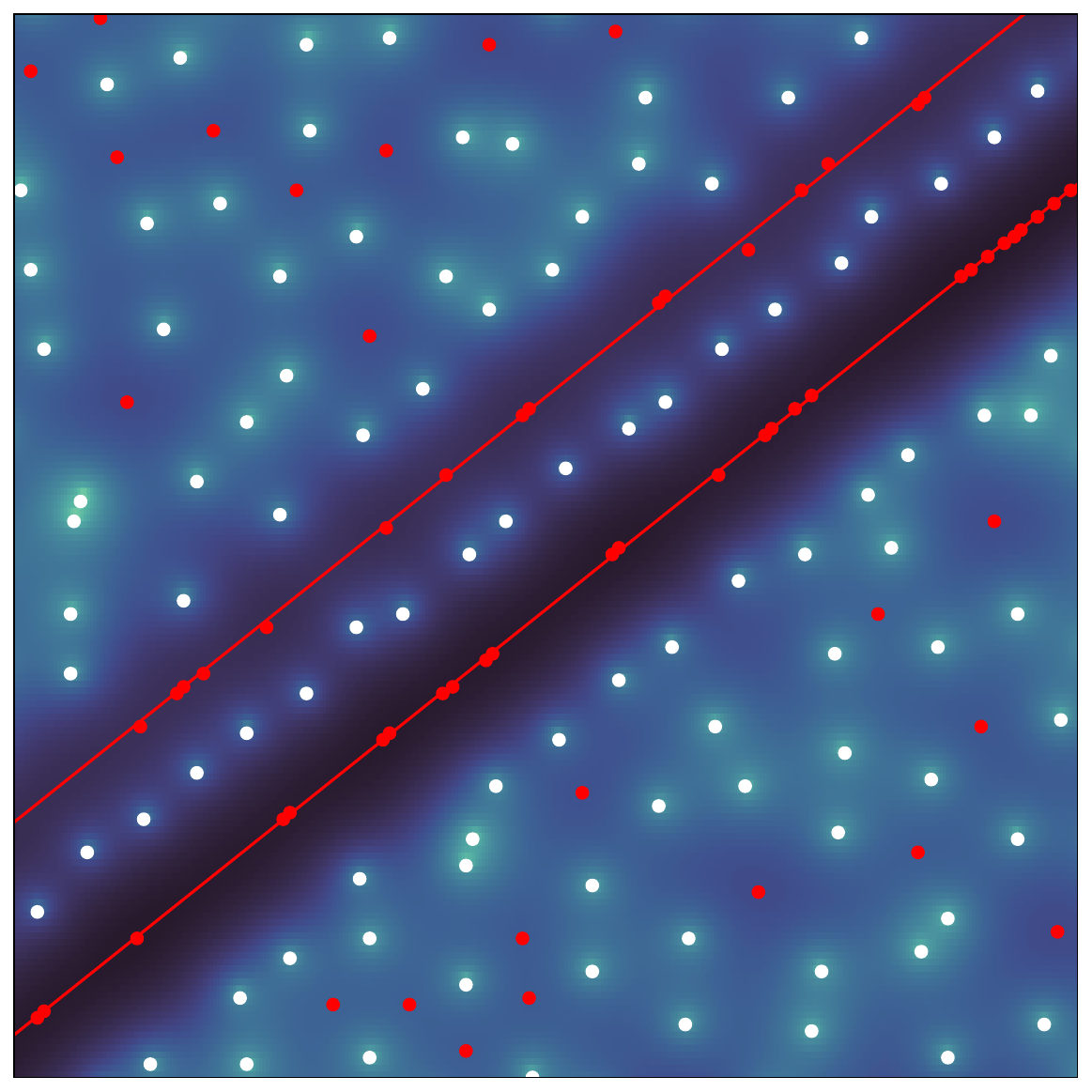}}
\hspace{1cm}
\subfloat{\includegraphics[width=3.5cm]{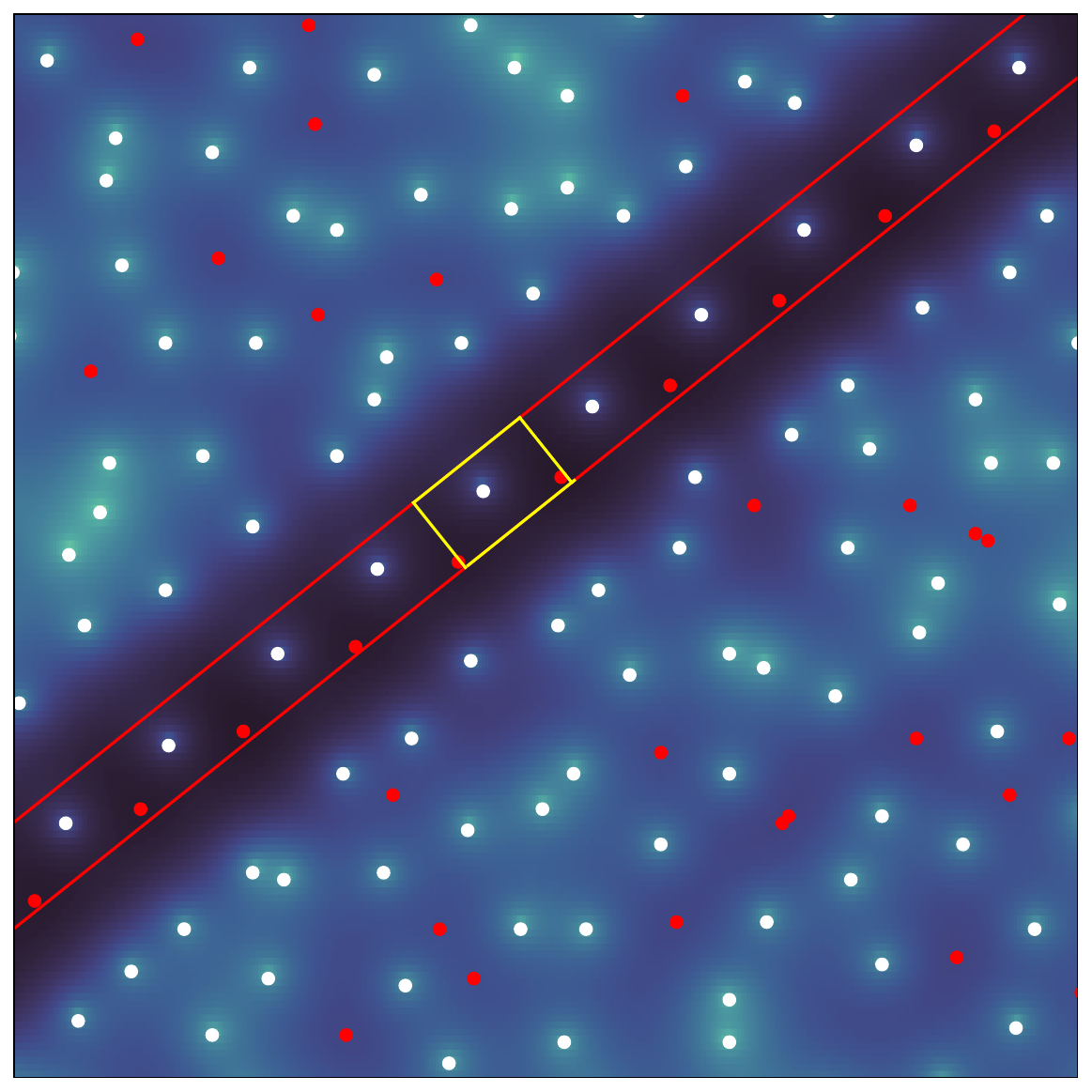}}
\hspace{1cm}
\subfloat{\includegraphics[width=3.5cm]{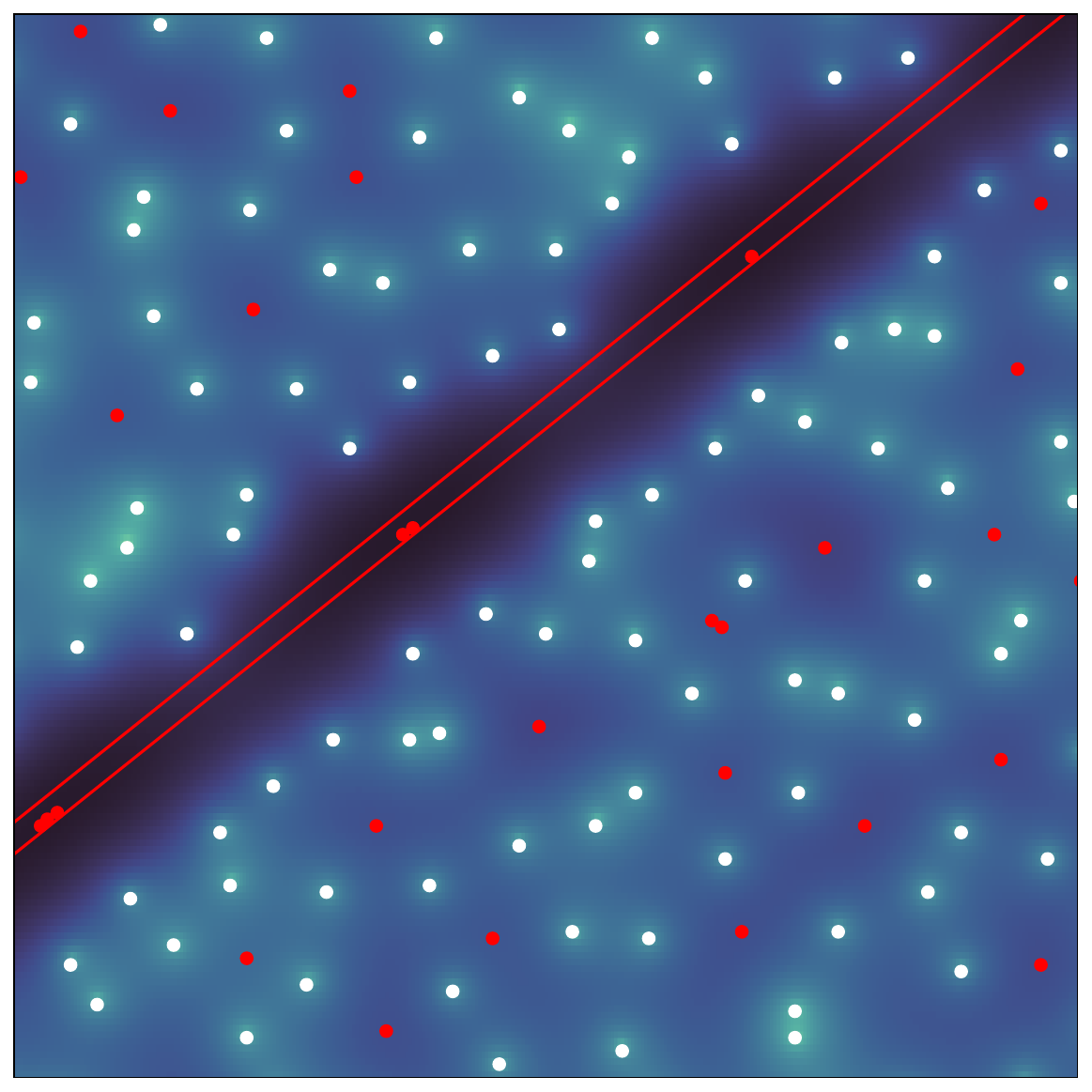}}\\
\subfloat{\includegraphics[width=3.5cm]{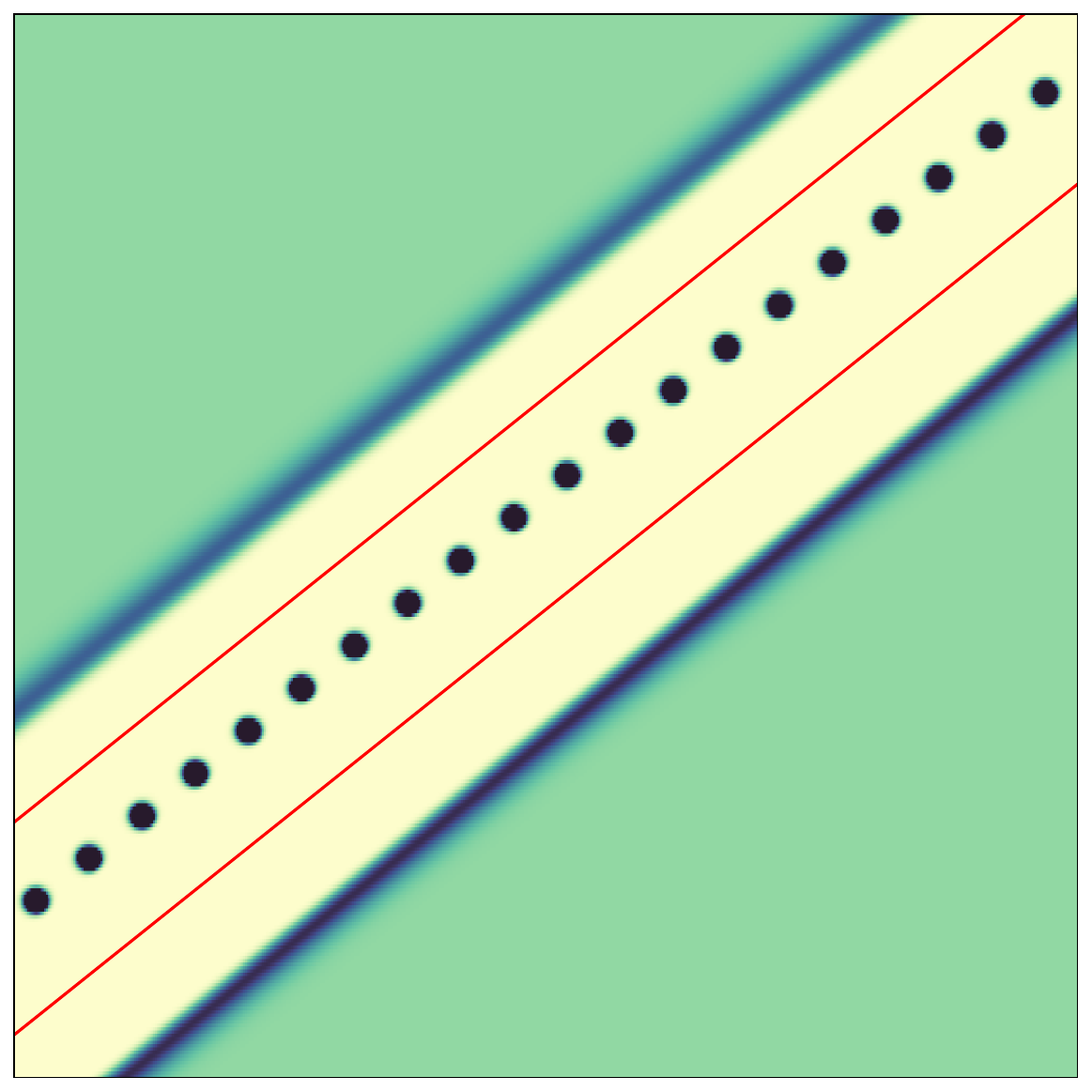}}
\hspace{1cm}
\subfloat{\includegraphics[width=3.5cm]{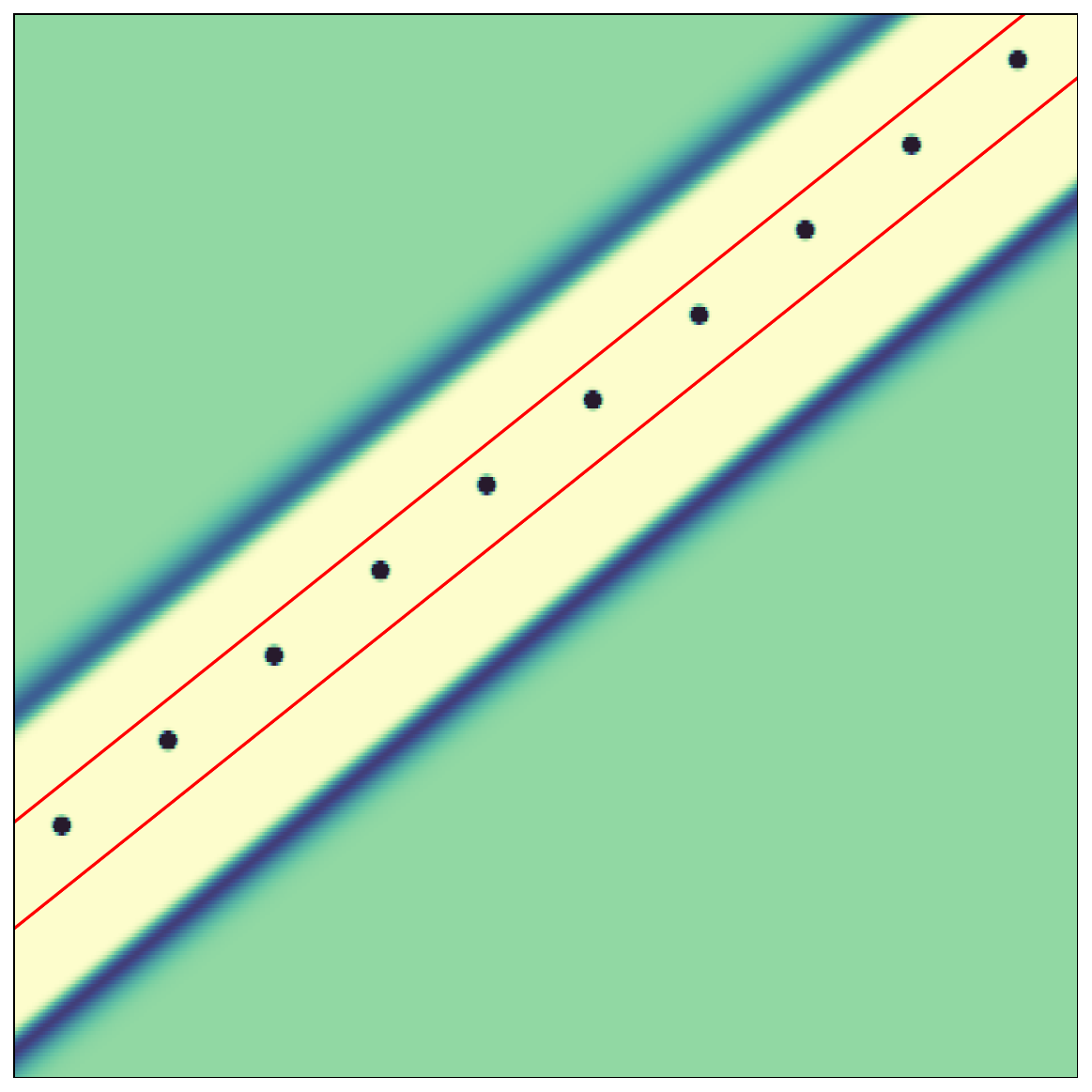}}
\hspace{1cm}
\subfloat{\includegraphics[width=3.5cm]{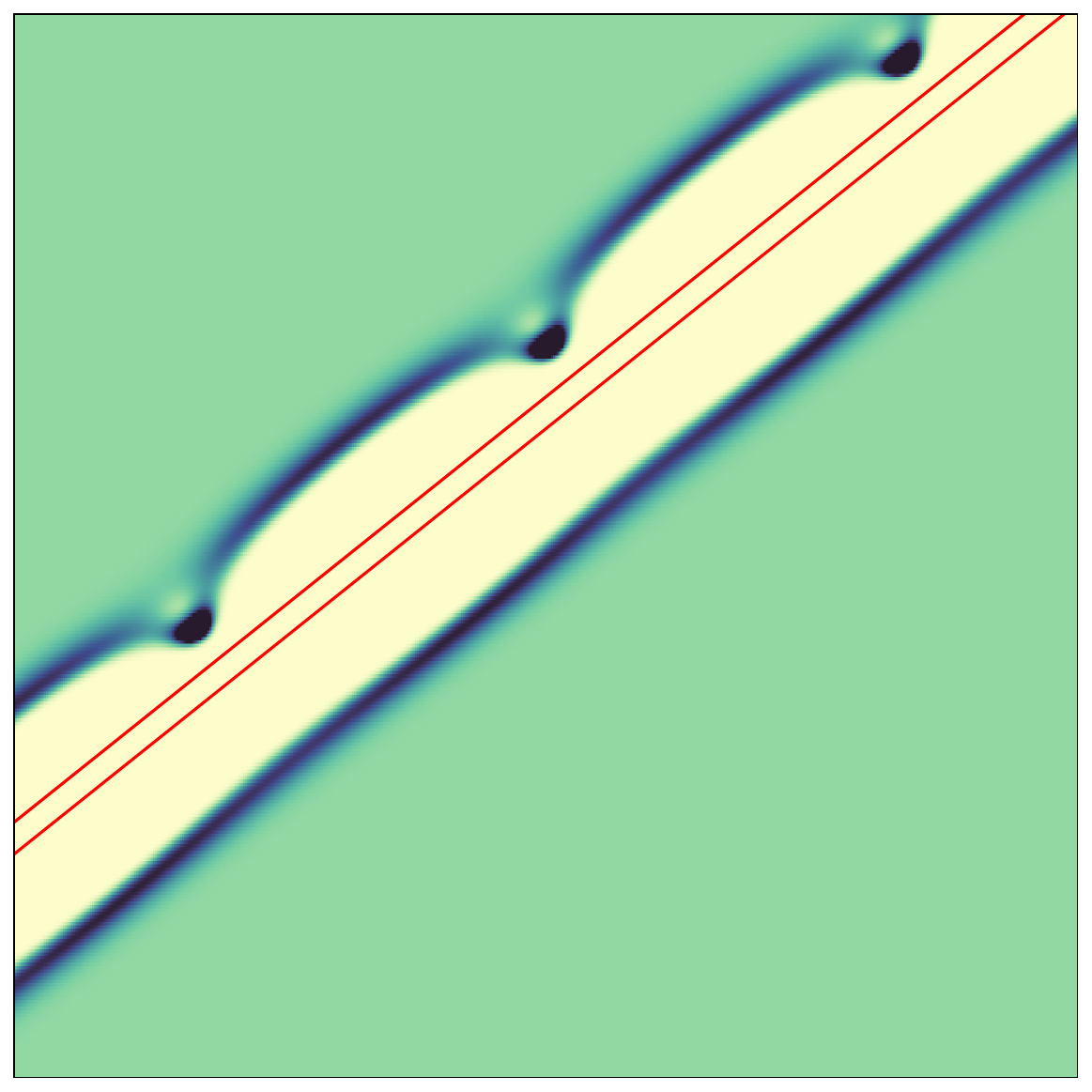}}
\caption{\small Spectrogram (top) of a noisy pair of chirps and corresponding density of zeros (bottom) for $\gamma_1=100$, $\gamma_2=40$, $a_1=-2$ (left), $-1$ (middle) or $-0.3$ (right), $a_2=0$ and $b=0.4$. 
Zeros are shown in white and local maxima in red. 
One of the rectangles $\cC_N$ in which a zero is trapped is shown in yellow.
\label{fig: Two_chirps}}
\end{figure}

\begin{prop}\label{prop: intensity_zeros_2chirps}
The intensity of the zeros of $\Spec(\xi+x)$ at $z\in\C$ only depends on the quantities $r$ and $s$ introduced in Figure~\ref{fig: Example_TwoChirps}, and
\begin{equation}\label{eq:intensity_zeros_2chirps}
\rho(r,s) = e^{-\Spec(x)(r,s)}\left(1+4\pi\sigma_b\left(\gamma_1r^2e^{-2\pi r^2}\right.\right. + \gamma_2(r-a)^2e^{-2\pi (r-a)^2}\left.\left. +2\sqrt{\gamma_1\gamma_2}r(r-a)e^{-\pi r^2-\pi (r-a)^2}\cos\left(2\pi as\right)\right)\right),
\end{equation}
where $\Spec(x)(r,s)$ corresponds to \eqref{eq:spectro_double_chirp}.
\end{prop}
\begin{proof}
Let $z\in\C$.
The expression \eqref{eq:intensity_zeros_2chirps} of the density of zeros can be obtained from \eqref{eq: rho_alternative} by computing $|\nabla_z' B(x)(\bar{z})|^2$, where $\nabla_{z}'$ is defined in \eqref{eq: def_nabla} and $x$ is defined as in \eqref{eq: signal_two_chirps}. We can expand $|\nabla_z' B(x)(\bar{z})|^2$ using the linearity of $\nabla_z'$ and the Bargmann transform into
\begin{equation}\label{eq: step1}
\gamma_1\left|\nabla_z' B(x_1)(\bar{z})\right|^2+\gamma_2\left|\nabla_z' B(x_2)(\bar{z})\right|^2+2\sqrt{\gamma_1\gamma_2}\Re\left(\nabla_z' B(x_1)(\bar{z})\overline{\nabla_z' B(x_2)(\bar{z})}\right).
\end{equation}
Identity \eqref{eq:Bargmann_chirp} leads to
$$\nabla_z' B(x_j)(\bar{z})=\frac{2(2b-i)\pi}{1+4b^2}(\omega-(a_j + 2b\tau))B(x_j)(\bar{z}),~j\in\{1,2\},$$
and therefore
\begin{equation}\label{eq: step2}
e^{-\pi|z|^2}\left|\nabla_z' B(x_j)(\bar{z})\right|^2=\frac{4\pi^2}{1+4b^2}|\omega-(a_j + 2b\tau)|^2\Spec(x_j)=\left\{
\begin{array}{l}
4\pi^2 r^2 \sigma_b e^{-2\pi r^2}~\mbox{if}~j=1;\\
4\pi^2 (r-a)^2 \sigma_b e^{-2\pi (r-a)^2}~\mbox{if}~j=2.
\end{array}\right.
\end{equation}
Moreover,
\begin{align}
\nabla_z' B(x_1)(\bar{z})\overline{\nabla_z' B(x_2)(\bar{z})}&=\frac{4\pi^2}{1+4b}(\omega-(a_1 + 2b\tau))(\omega-(a_2 + 2b\tau))B(x_1)(\bar{z})\overline{ B(x_2)(\bar{z})}\nonumber\\
&=4\pi^2 r(r-a)B(x_1)(\bar{z})\overline{ B(x_2)(\bar{z})}.\label{eq: step3}
\end{align}
Finally, combining \eqref{eq: step3} with \eqref{eq: cross_terms} and injecting it, along with \eqref{eq: step2}, into \eqref{eq: step1} gives \eqref{eq:intensity_zeros_2chirps}.
\end{proof}

As can be observed in Figures \ref{fig: Example_TwoChirps} and \ref{fig: Two_chirps}, the spectrogram of the noisy pair of chirps preserves the same strip of zeros as the noiseless spectrogram. 
This is a consequence of Rouché's theorem, combined with the fact that when the two chirps are close enough, then each zero of the noiseless spectrogram is surrounded by high values of the spectrogram \citep{Meignen}. 
More precisely, we define the rectangle $\cC_N$, $N\in\N$, as the boundary of the points such that $(r,s)\in [0, a]\times [N/a, (N+1)/a]$; see Figure \ref{fig: Example_TwoChirps}.

\begin{prop} \label{prop: zero_trapped_general}
Let $\epsilon\in (0, 1/4)$ and define $M_{\cC_N}$ as in Proposition \ref{prop: Gaussian_bound}. 
Assume
\begin{enumerate}
\item $|\log(\gamma_1) - \log(\gamma_2)|< 2\pi a^2$; \label{ass: 1}
\item $a\leq \sqrt\frac{2}{\pi}~\textnormal{or}~\frac{1}{2}|\log(\gamma_1)-\log(\gamma_2)|\geq -\arccosh\left(\pi a^2 - 1\right)+a\sqrt{\pi^2 a^2-2\pi}$; \label{ass: 2}
\item $\min\left(\left|\sqrt\gamma_1-\sqrt\gamma_2e^{-\pi a^2}\right|,\left|\sqrt\gamma_2-\sqrt\gamma_1e^{-\pi a^2}\right|\right) \geq \frac{2}{\sqrt{\sigma_b}}\left( M_{\cC_N} + \sqrt{\log(4/\epsilon)}\right),$ \label{ass: 3}
\end{enumerate}
Then, for each $N\in\N$, $\Spec(\xi+x)$ has a unique zero inside $\cC_N$ with probability at least $1-\epsilon$.
\end{prop}

\begin{proof}
By Proposition \ref{prop: location_zeros}, $\Spec(x)$ has a unique simple zero in the interior of each rectangle $\cC_N$ when $|\log(\gamma_1) - \log(\gamma_2)|< 2\pi a^2$. Using Proposition \ref{prop: Rouche}, we show that this unique zero inside $\cC_N$ stays trapped when adding noise under appropriate assumptions. We first split $\cC_N$ into $3$ sections and give a lower bound for $\Spec(x)$ on each one:
\begin{itemize}
\item \underline{If $r=0$ and $s\in[N/a, (N+1)/a]$}:
\begin{align*}
\Spec(x)(r,s)&=\sigma_b\left(\gamma_1+\gamma_2e^{-2\pi a^2}+2\sqrt{\gamma_1\gamma_2} e^{-\pi a^2}\cos\left(2\pi as\right)\right)\geq \sigma_b\left(\sqrt\gamma_1-\sqrt\gamma_2e^{-\pi a^2}\right)^2.
\end{align*}

\item \underline{If $r=a$ and $s\in[N/a, (N+1)/a]$}:
\begin{align*}
\Spec(x)(r,s)&=\sigma_b\left(\gamma_1e^{-2\pi a^2}+\gamma_2+2\sqrt{\gamma_1\gamma_2} e^{-\pi a^2}\cos\left(2\pi as\right)\right)\geq \sigma_b\left(\sqrt\gamma_2-\sqrt\gamma_1e^{-\pi a^2}\right)^2.
\end{align*}

\item \underline{If $r\in[0, a]$ and $s=N/a$ or $s=(N+1)/a$}:
$$\Spec(x)(r,s)=\sigma_b\left(\sqrt\gamma_1e^{-\pi r^2}+\sqrt\gamma_2e^{-\pi (r-a)^2}\right)^2.$$
It was shown in \cite[Appendix A]{Meignen} (where their notations $\gamma, \alpha$ and $A$ corresponds in our setting to, respectively, $1$, $\sqrt{\frac{\pi}{2}}a$ and $\sqrt{\frac{\gamma_2}{\gamma_1}}$) that under Assumption \eqref{ass: 2} the two chirps are close enough so that the function
$r\mapsto \sqrt\gamma_1e^{-\pi r^2}+\sqrt\gamma_2e^{-\pi (r-a)^2}$ increases until it reaches a unique local maxima and then decreases. Therefore, $\Spec(x)$ takes its lowest value either for $r=0$ or $r=a$ so it is lower bounded by
$$\sigma_b\min\left(\sqrt\gamma_1+\sqrt\gamma_2e^{-\pi a^2},\sqrt\gamma_2+\sqrt\gamma_1e^{-\pi a^2}\right)^2\geq \sigma_b\min\left(\left(\sqrt\gamma_1-\sqrt\gamma_2e^{-\pi a^2}\right)^2,\left(\sqrt\gamma_2-\sqrt\gamma_1e^{-\pi a^2}\right)^2\right).$$
\end{itemize}

\noindent Combining these three cases gives
\begin{equation}\label{eq: first_bound}
\forall z\in \cC_N,~\Spec(x)(z)\geq \sigma_b\min\left(\left(\sqrt\gamma_1-\sqrt\gamma_2e^{-\pi a^2}\right)^2,\left(\sqrt\gamma_2-\sqrt\gamma_1e^{-\pi a^2}\right)^2\right).
\end{equation}
Finally, using \eqref{ass: 3} and Propositions \ref{prop: Rouche} and \ref{prop: Gaussian_bound}, we can bound the probability of observing exactly one point inside $\cC_N$ by
\begin{equation*}
1-4\exp\left(-\left(\frac{\sqrt{\sigma_b}}{2}\min\left(\left|\sqrt\gamma_1-\sqrt\gamma_2e^{-\pi a^2}\right|,\left|\sqrt\gamma_2-\sqrt\gamma_1e^{-\pi a^2}\right|\right) - M_{\cC_N}\right)^2\right)\geq 1 - \epsilon.\qedhere
\end{equation*}
\end{proof}

We note that for two chirps of equal strength ($\gamma_1=\gamma_2=\gamma$), Assumption \eqref{ass: 1} is trivially satisfied, Assumption \eqref{ass: 2} simplifies to $a\geq \sqrt{2/\pi}$, and Assumption \eqref{ass: 3} simplifies to 
$$
   \gamma \geq \frac{-4\log(\epsilon/4)}{\sigma_b\left(1-e^{-\pi a^2}\right)^2}.
$$
So if the chirps are close enough and strong enough then we are very likely to see a zero trapped in each rectangle $\cC_N$. To wit, when $\gamma_1\neq \gamma_2$, the assumptions of \eqref{prop: zero_trapped_general} are harder to interpret, but shows that if the chirps are close to each other, the strongest chirp can push the line \eqref{eq:location_zeros} of zeros \emph{out of} the region between the chirps, to the side of the chirp with lower amplitude; see Figure \ref{fig: Two_chirps}. 
Figure \ref{fig: Two_chirps} also shows that this line of zeros can be preserved when the inter-chirp distance is close to zero, while the local maxima of both chirps have fused into a single strip.
This stark difference hints at zeros being a better tool at detecting the presence of interference. 
For better visualization, we provide as supplementary material and with the associated code an animation showing the behavior of zeros and maxima as two chirps with different amplitude get close to each other.

\section{Discussion}
Through explicit computation of the intensity \eqref{eq:zero_intens_signal} of the point process of zeros and direct applications of Rouché's theorem, we illustrated how zeros tend to delineate the signal support and why they can be trapped. 
In terms of applications, zeros arising from interference may thus be preferable to maxima when trying, e.g., to count superimposed simple components.
   Furthermore, Identity \eqref{eq:zero_intens_signal} also opens the door to using zero locations for signal reconstruction in a parametric setting. 
   As seen in the case of Hermite functions, single chirps and multiple chirps, we can obtain an exact expression of the relationship between the signal parameters and the average number of zeros falling inside a given boundary, like in \eqref{eq: Expect_ball_hermite} and \eqref{e:expected_number}. 
   Adding the knowledge of the location of trapped zeros, these results could be used to construct statistics based on zero locations to estimate signal parameters.

\medskip

One limitation of the setting in this paper is the restriction to Gaussian noise.
Indeed, we rely on properties of Gaussian random fields, namely the Kac-Rice formula and tail inequalities for the maximum of Gaussian random fields. This is a common restriction when studying the location of critical points of random fields; see e.g. \cite{Abreu, BaFlCh18, MACLM24}. 
While tail inequalities exist for generic random fields and the Kac-Rice formula also holds for any random field that is regular enough, see \cite[Theorem 6.7]{Azais}, the latter is challenging to use in practice due to the difficulty of computing conditional expectations like \eqref{eq: Cond_Exp_KR} for non-Gaussian fields.

\medskip

Finally,  natural extensions of this work would be to characterize the process of local maxima, as initiated by \cite{Abreu} for white noise, and to consider non-analytic STFTs that come with non-Gaussian windows \citep{HaKoRo22}.

\appendix

\section{Technical results}

\subsection{Domination of the noiseless signal} \label{sec: Gaussian_bound}

\begin{prop}\label{prop: Gaussian_bound}
   Let $\cC$ be a closed, simple curve of $\C$. 
   Let $x:\R\rightarrow\C$ be a locally integrable function, and let $\xi$ be the white Gaussian noise introduced in Section \ref{s:main_results}.
   Define $M_{\cC}=\E[\sup_{z\in \cC} F(z)]$, where $F$ is a real Gaussian process on $\bC$ with kernel
   $$
      \kappa(z, w)=\frac{1}{2}e^{-\frac{\pi}{2}|z-w|^2}\cos(\pi\Im(z\bar w)).
   $$
   If $\inf_{z\in \cC} \Spec(x)(z) > 2M_{\cC}^2$ then,
   $$\mathbb{P}\left(\forall z\in \cC,~\Spec(\xi)(z) < \Spec(x)(z)\right) \geq 1-4\exp\left(-\left(\sqrt{\frac{\inf_{z\in \cC} \Spec(x)(z)}{2}} - M_{\cC}\right)^2\right)$$
\end{prop}

\begin{proof}
Both $z\mapsto e^{-\pi|z|^2/2}\Re(B(\xi)(z))$ and $z\mapsto e^{-\pi|z|^2/2}\Im(B(\xi)(z))$ are real centered Gaussian processes on $\C$ with kernel
$$\frac{1}{2}\Re(e^{\pi z \bar w})e^{-\pi|z|^2/2}e^{-\pi|w|^2/2} = \kappa(z, w)$$
satisfying $\kappa(z, z)=1/2$ for all $z\in\C$ and $\kappa(z, w) = \kappa(\bar z, \bar w)$.
In particular, for $z\in\C$, $F(z)$ and $F(\bar z)$ have the same distribution. 
Then
$$
   M_{\cC} = M_{\bar \cC} = \E\left[e^{-\pi|z|^2/2}\sup_{z\in \bar \cC}\Re(B(\xi)(z))\right] = \E\left[e^{-\pi|z|^2/2}\sup_{z\in \bar \cC}\Im(B(\xi)(z))\right].
$$
Applying classical tail inequalities for the supremum of real-valued Gaussian fields (see for instance \cite[Identity (2.31)]{Azais}) yields, for all $u \geq M_{\cC}$,
\begin{equation}\label{eq: gaussian_sup_bound}
   \mathbb{P}\left(\sup_{z\in \bar \cC}e^{-\pi|z|^2/2}\Re(B(\xi)(z))>u\right) \leq \exp\left(-\frac{(u - M_{\cC})^2}{2\sup_{z\in \bar \cC}\kappa(z, z)}\right) = \exp\left(-(u - M_{\cC})^2\right),
\end{equation}
with the same inequality also holding for $\Im(B(\xi))$. Therefore, for any $u\geq M_{\cC}$,
\begin{align*}
\mathbb{P}\left(\sup_{z\in \cC} \Spec(\xi)(z) > u\right)
&\leq \mathbb{P}\left(\sup_{z\in \bar \cC} e^{-\pi|z|^2/2}|\Re(B(\xi)(z))| > \sqrt{\frac{u}{2}}\right) + \mathbb{P}\left(\sup_{z\in \bar \cC} e^{-\pi|z|^2/2}|\Im(B(\xi)(z))| > \sqrt{\frac{u}{2}}\right)\\
&\leq 4\exp\left(-\left(\sqrt{\frac{u}{2}} - M_{\cC}\right)^2\right).
\end{align*}
In particular, if $\inf_{x\in \cC} \Spec(x)(z) > 2M_{\cC}^2$,
\begin{equation*}
\mathbb{P}\left(\forall z\in \cC,~\Spec(\xi)(z) < \Spec(x)(z)\right) \geq 1-4\exp\left(-\left(\sqrt{\frac{\inf_{x\in \cC} \Spec(x)(z)}{2}} - M_{\cC}\right)^2\right)\qedhere
\end{equation*}
\end{proof}

\subsection{Expression of the spectrogram of two parallel chirps} \label{sec: Spectro_two_chirps}

For the linear chirp $x_1:t \mapsto e^{2i\pi t(a_1+bt)}$, we have, for $z\in \C$, 
\begin{align}
   B(x_1)(z) &= 2^{1/4}\int_{\R} e^{(2i\pi b -\pi)t^2 + (2i\pi a_1 + 2\pi z)t -\frac{\pi}{2}z^2}\der t\nonumber\\
   &= 2^{1/4}\exp\left((2i\pi b -\pi)\left(\frac{ia_1 + z}{2i b - 1}\right)^2 -\frac{\pi}{2}z^2\right)\int_{\R}
   \exp\left((2i\pi b -\pi)\left(t + \frac{ia_1 + z}{2i b - 1}\right)^2\right)\der t\nonumber\\
   &= \frac{2^{1/4}}{\sqrt{1-2ib}}\exp\left(-\pi\frac{(ia_1 + z)^2}{2i b - 1} - \frac{\pi}{2}z^2 \right).\label{eq:Bargmann_chirp}
\end{align}
Replacing $z$ by $\tau + i\omega$ and simplifying the expression leads to
\begin{equation}\label{eq: Spectro_chirp}
\Spec(x_1)(z) = e^{-\pi|z|^2}\left|B(x_1)(\bar{z})\right|^2=\sqrt{\frac{2}{1+4b^2}} \exp\left(-\frac{2\pi}{1+4b^2}(\omega-(a_1+2b\tau))^2 \right)
\end{equation}
and thus to \eqref{e:spectrogram_single_chirp} after injecting the expression of $\sigma_b$ and $r$ into \eqref{eq: Spectro_chirp}. 
We now consider the signal
\begin{equation}\label{eq: signal_two_chirps}
   x:t \mapsto \sqrt{\gamma_1}x_1(t)+ \sqrt{\gamma_2}x_2(t)~\mbox{with}~x_j(t)=e^{2i\pi t(a_j+bt)}, j\in\{1, 2\},
\end{equation}
whose spectrogram writes, for $z\in\C$, 
\begin{equation}\label{eq: Spectro_double_chirp_expansion}
   \Spec(x)(z)=\gamma_1\Spec(x_1)(z)+\gamma_2\Spec(x_2)(z)+2\sqrt{\gamma_1\gamma_2}e^{-\pi|z|^2}\Re\left(B(x_1)(\bar{z})\overline{B(x_2)(\bar{z})}\right).
\end{equation}
Identity \eqref{eq:Bargmann_chirp} with some straightforward computations leads to $e^{-\pi|z|^2}\Re\left(B(x_1)(\bar{z})\overline{B(x_2)(\bar{z})}\right)$ being equal to
\begin{equation}\label{eq: cross_terms}
\sigma_b\exp\left(-\pi\frac{\sigma_b^2}{2}(\omega-(a_2+2b\tau))^2-\pi\frac{\sigma_b^2}{2}(\omega-(a_1+2b\tau))^2\right)\\
\times \cos\left(\pi\sigma_b^2(a_2-a_1)\left(\tau+2b\omega-(a_1+a_2)b\right)\right).
\end{equation}
Injecting the expressions of $\sigma_b$, $r$ and $s$ into \eqref{eq: cross_terms}, combined with \eqref{eq: Spectro_double_chirp_expansion} and \eqref{e:spectrogram_single_chirp}, gives \eqref{eq:spectro_double_chirp}.






\singlespacing
\setlength{\bibsep}{1.8pt}

\bibliographystyle{elsarticle-harv}
\bibliography{ref, stats, stft}

\end{document}